\documentclass[a4paper,11pt]{article}
\usepackage{amsmath,amssymb,cite}
\usepackage{graphicx,amssymb,bm,latexsym,color,epsf}
\usepackage{soul}
\usepackage{hyperref} 
\hypersetup{colorlinks=true, citecolor=blue, filecolor=black, linkcolor=blue, urlcolor=blue, pdfpagemode=UseNone}

\makeatletter
\@addtoreset{equation}{section}

\makeatother
\newcommand{\be}{\begin{equation}}
\newcommand{\ee}{\end{equation}}
\newcommand{\bea}{\begin{eqnarray}}
\newcommand{\eea}{\end{eqnarray}}
\newcommand{\ena}{\end{eqnarray}}
\newcommand{\vs}[1]{\vspace{#1 mm}}

\newcommand{\cA}{{\cal A}}
\newcommand{\cR}{{\cal R}}

\newcommand{\cO}{{\cal O}}

\newcommand{\Tr}{{\rm Tr}}

\newcommand\ie{\textit{i.e.}\ }
\newcommand\eg{\textit{e.g.}\ }
\newcommand\cf{\textit{cf.}\ }

\newcommand{\half}{\tfrac{1}{2}}

\begin{document}

\begin{center}
{\Large\bf Trace anomaly and infrared cutoffs}
\vs{6}

{\large
Tim R. Morris\footnote{e-mail address: T.R.Morris@soton.ac.uk}$^{a}$
and
Roberto Percacci\footnote{e-mail address: percacci@sissa.it}$^{,b,c}$
} \\
\vs{6}

$^a${\em STAG Research Centre \& Department of Physics and Astronomy,\\
University of Southampton, Highfield, Southampton, SO17 1BJ, U.K.}

$^b${\em International School for Advanced Studies, via Bonomea 265, I-34136 Trieste, Italy}

$^c${\em INFN, Sezione di Trieste, Italy}


\end{center}
{\bf Abstract.}
The Effective Average Action is a form of effective action
which depends on a cutoff scale suppressing the contribution
of low momentum modes in the functional integral.
It reduces to the ordinary effective action when the cutoff
scale goes to zero.
We derive the modifications of the scale Ward identity due to this cutoff and show how the resulting identity then intimately relates the trace anomaly to the Wilsonian realisation of the renormalization group.


\section{Introduction}

A theory that is scale invariant at the classical level, is in general no longer scale invariant at the quantum level. 
The breaking of scale invariance is known as the trace anomaly. It has two causes. On the one hand a non-flat external metric leads to contributions proportional to integrals of curvatures of the metric \cite{Capper:1974ic,Duff:1993wm}. On the other hand, there is a part proportional to the beta functions of the theory \cite{coleman,Chanowitz:1972da,Adler:1976zt,Collins:1976yq}. The first type of contribution occurs even for a free field theory, while the second type of contribution appears even in flat space. In this paper we will be concerned almost exclusively with this last situation.

A scale transformation is a change of all lengths by a 
constant factor.
One can interpret this either as a rescaling of the coordinates, or as a rescaling of the metric (see Appendix A).
Even though we will not deal with gravity in this paper,
we choose the latter interpretation. 
For simplicity we will deal mostly with a single scalar field $\phi$.
The infinitesimal transformation of the fields is then
\bea
\delta_\epsilon g_{\mu\nu}&=&2\epsilon g_{\mu\nu}
\nonumber\\
\delta_\epsilon \phi&=&\epsilon d_\phi \phi
\label{scaletr}
\eea
where 
$d_\phi=-\frac{d-2}{2}$
is the canonical length dimension of $\phi$ in $d$ spacetime dimensions. We work in Euclidean signature 
where the energy-momentum tensor is defined by
\be 
T^{\mu\nu}=-\frac{2}{\sqrt g}\frac{\delta S}{\delta g_{\mu\nu}}\,.
\label{emt}
\ee
In concordance with the quantum action principle \cite{Schwinger:1951xk}, under a scale transformation the operator $ -\delta_\epsilon S$ is inserted into correlation functions, where $S$ is the bare action. If couplings in the theory are dimensionful, already at the classical level there will be a breaking of scale invariance through this contribution. We are interested in the case that the theory is scale invariant at the classical level, so we will have dimensionless couplings only. For the theory of a single scalar field in $d=4$ spacetime dimensions, that means the interaction potential will just be
\be 
\label{phi4}
V(\phi) = \frac{\lambda}{4!}\,\phi^4\,,
\ee
where $\lambda$ is the dimensionless coupling.
When everything is written in renormalized terms, the result is then the insertion of the renormalised operator
\be 
\label{breaking}
\epsilon\int_x T^\mu{}_\mu= \epsilon\,\beta\int_x \frac{1}{4!}\phi^4\ ,
\ee
where $\beta = \mu \partial_\mu \lambda(\mu)$ is the 
Renormalization Group (RG) beta function
and we denote $\int_x=\int d^dx\sqrt{g}$ the integration over spacetime.  
In particular, for the Legendre Effective Action (EA) (``0-point function'')
\be 
\label{deltaGamma1}
\delta_\epsilon \Gamma 
= -\cA(\epsilon)
\ ,
\ee
where $\cA(\epsilon)$ is the expectation value of \eqref{breaking} in the presence of sources,
and is called the ``trace anomaly''.
\footnote{This anomaly is present also in curved space,
as demonstrated for a spherical background in \cite{Drummond:1977dg}
and (using the background field method) on an arbitrary
background in \cite{Alves:1987zm,BoschiFilho:1991xz}.}
Equation (\ref{breaking}) then shows that the anomaly
vanishes, and scale invariance is realized also
in the quantum theory, at a fixed point.

Evidently (\ref{breaking}) follows only if the breaking of scale invariance is solely due to the running at the quantum level of the dimensionless coupling $\lambda$, rather than any other mass scale introduced into the theory. 
It therefore depends upon details of the regularization and renormalization procedure.
For example in dimensional regularisation, scale invariance is broken by the fact that the bare coupling $\lambda_0$ is dimensionful outside four dimensions, where it is rewritten as $\mu^{d-4}$ times a series in the renormalized coupling $\lambda$ and $1/(d-4)$.
This results only in the term (\ref{breaking}). 
With a momentum cutoff or other physical (dimensionful) regulator, one encounters a quadratic divergence and the renormalized mass (defined as the second derivative of the potential at zero field) then becomes an arbitrary parameter. This would give rise to additional terms proportional to $\int_x\phi^2$ in the r.h.s. of (\ref{breaking}), 
with a scheme-dependent coefficient.
\footnote{We call this an anomaly because the classical action is invariant whereas the quantum action is not. In the literature it is more standard to call anomaly only the breaking of a symmetry that cannot be fixed by a local counterterm,
as is the case for the term (\ref{breaking}).}
Among all these renormalized theories one can look for the one with the least breaking of scale invariance: this is the ``critical'' theory where the renormalized mass is exactly zero. Still, as we shall discuss in Section 2, scale invariance is broken by exactly the same anomaly (\ref{breaking}) as in dimensional regularization.
This is a physical effect that cannot be removed by
renormalization or improvements to the bare action. Scale invariance is present at the classical level, but at the quantum level it is broken by an irreducible amount.
We may say that the critical theory ``almost'' realizes  scale invariance at the quantum level.

To see what (\ref{breaking}) implies in the critical theory, we note that inserting $\frac1{4!}\int_x \phi^4$ is achieved by differentiating with respect to $\lambda$.  Recalling the extra sign in ${\rm e}^{ -\Gamma}$, we should therefore expect
in general,
\be 
\label{deltaGamma2}
\cA(\epsilon) 
= \epsilon\,\beta(\lambda)\, \partial_\lambda \Gamma\,.
\ee
The signs in (\ref{deltaGamma1},\ref{deltaGamma2}) can be understood when we recall that scale transformations \eqref{scaletr} increase length scales for positive $\epsilon$, and thus decrease mass scales, \ie
associate $-\epsilon$ with unit positive mass dimension, 
as we see from \eqref{scaletr}. In other words, we can think of $\delta_\epsilon$ as generating a flow towards the infrared.

It is an old idea that mass scales in nature may be
of quantum mechanical origin, as is indeed true to a large extent in QCD.
For a scalar theory this is related to 
the Coleman-Weinberg potential \cite{Coleman:1973jx}.
This idea has seen a revival in recent years
\cite{Meissner:2007xv,Hill:2014mqa,Gies:2015lia,Wetterich:2019qzx}, 
see also 
\cite{Wetterich:1987fm,Salvio:2014soa,Einhorn:2014gfa,Ferreira:2016wem,Oda:2018zth}
for similar ideas in a cosmological and gravitational context.
In this paper we will explore the implications of classical
scale invariance in the context of the Effective Average Action (EAA)
$\Gamma_k$,
which is a version of the EA supplied with an infrared cutoff $k$,
and reducing to the EA when $k\to0$
\cite{Wetterich:1992yh,Morris:1993qb}.
Our main result is the following:
when the classical action is scale invariant, in addition to the RG flow for the EAA, 
there is a Ward Identity (WI) for scale transformations
which takes the form
\be
\delta_\epsilon \Gamma_k=-\cA(\epsilon)
+\epsilon\partial_t\Gamma_k\,,
\label{wi}
\ee
where the second term represents 
the RG flow due to $k$ ($t=\log k$).
In the rest of the paper we demonstrate in detail, 
using momentum cutoffs, how this anomaly
arises and how it reduces to (\ref{deltaGamma1}) in the limit
$k\to0$.
There is partial overlap with the earlier work of 
\cite{Delamotte:2015aaa}, who also considered the effect of the Wilsonian
cutoff on scale transformations, but did not take
into account the UV contributions to the anomaly,
because they were not relevant to their problem.

The paper is organized as follows.
In section 2 we discuss the trace anomaly
in the context of theories with momentum cutoffs:
either UV, or IR or both.
In section 3 we derive the WI (\ref{wi})
and recall how in some circumstances
it can be applied to gain information on the EA.
Section 4 deals with the form of the anomaly and the EAA
in approximate treatment. We consider the one loop
approximation and other popular approximations
such as the derivative expansion or the vertex expansion.
Section 5 is devoted specifically to the derivation of 
(\ref{deltaGamma1}) from (\ref{wi}) in the limit $k\to0$.
In section 6 we briefly discuss the realization of
quantum scale invariance at fixed points and
in section 7 we make some connections to other
ideas in the literature and point to some
possible developments.

\section{UV  cutoff and the trace anomaly}
\label{sec:Feynman}

\begin{figure}[ht]
\centering
\includegraphics[scale=0.35]{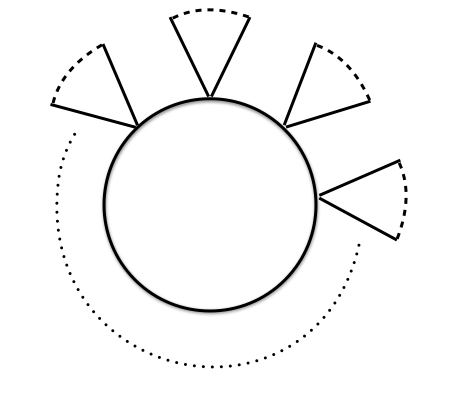}
\caption{Feynman diagrams contributing to the effective action.}
\label{fig:Feynman}
\end{figure}

In order to develop some intuition for the workings
of the anomaly when using momentum cutoff as regularization,
it will be helpful to start from a perturbative treatment
based on standard diagrammatic methods.
We will consider the effect of both UV and IR cutoffs. If we work to one loop we can write the bare action as 
\be 
S[\phi] = \int_x \left[ \half(\partial_\mu\phi)^2 + V(\phi)\right]\,,
\ee
for some potential $V(\phi)$.
Expanding the one loop EA, we can write
\be 
\label{sum}
\Gamma[\phi] = S[\phi] +\sum_{n=1}^\infty \mathcal{V}_n[\phi]
\ee
where $\phi$ denotes here, 
by a slight abuse of language,
the classical VEV of the corresponding quantum field,
\be
\mathcal{V}_n[\phi]=-\frac12\frac{(-1)^n}{n}\Tr \left(\frac{1}{-\partial^2}\, V''\right)^n\,,
\ee
and we have thrown away the field independent part.
This is a sum over the Feynman diagrams as indicated in fig. \ref{fig:Feynman}. 
One finds
\be 
\label{vertex}
\mathcal{V}_n[\phi] = -\frac{(-1)^n}{2n}  \int_{p_i,\cdots,p_n}\!\!\!\!\!\!\!\!\!\!\!\!\!\! V''(p_1)\cdots V''(p_n)\, (2\pi)^d\delta(p_1+\cdots+p_n)\, A(p_1,\cdots,p_n)\,.
\ee
The unconstrained momentum integral for each  diagram takes the form 
\be 
\label{A}
A(p_1,\cdots, p_n) =\int_q \frac{1}{q^2 (q+P_1)^2 (q+P_2)^2 \cdots (q+P_{n-1})^2}\,,
\ee
where $P_j = \sum_{i=1}^j p_j$  ($P_n=0$ being enforced by the momentum conserving $\delta$-function) are partial sums of the external momenta injected into the diagram by 
\be 
\label{insertion}
V''(p_i)=\int_x V''\left(\phi(x)\right) \, {\rm e}^{ip_i\cdot x}\,,
\ee
and in the integrals over momenta we include the usual factor of $(2\pi)^{-d}$. The integrals \eqref{A} are infrared finite provided we choose non-exceptional external momenta, \ie provided that  $\sum_{i=j}^k p_i \ne 0$ for all $1\!\le\! j\!<\!k\!\le\! n$.\footnote{In the case that this is violated we have in the denominator (at least) one $(q+P_j)^4$ term which is (at least)  logarithmically IR divergent.} Furthermore for $n\!>\!2$, 
these Feynman diagrams are  ultraviolet finite. For a field $\phi(p)$ with some suitable smooth behaviour in momentum space, we can therefore define these $n>2$ contributions rigorously. Thus provided that the limiting behaviour of $A(p_1,\cdots, p_n)$ as momenta become exceptional, is still integrable when the complete vertex is considered, the $\mathcal{V}_{n>2}[\phi]$ are well defined.

Working in four dimensions, the insertions \eqref{insertion} have mass dimension -2. Taking into account all the other parts in \eqref{vertex}, it is straightforward to verify that the $\mathcal{V}_{n}[\phi]$  are dimensionless, as they must be to be part of $\Gamma$. If further we use the potential \eqref{phi4}, then no dimensionful coupling is included. Since the $n\!>\!2$ contributions do not need
regularisation, it follows that they are scale invariant, \ie vanish under  the operation $\delta_\epsilon$. 

It will be helpful to show this in detail however, since we will then need to break the invariance with a cutoff. Recall that the scale variation is actually being carried by the fields and the metric, as in eqn. \eqref{scaletr}. Our metric is currently flat: $g_{\mu\nu}=\delta_{\mu\nu}$, but its inverse is present in $\partial^2=g^{\mu\nu}\partial_\mu\partial_\nu$, which means that  its eigenvalues transform as 
\be 
\label{q2}
\delta_\epsilon q^2 = -2\epsilon q^2\,.
\ee 
Since we should thus regard momentum as having a lower index, $p\cdot x$ does not contain the metric, and therefore the Fourier transform \eqref{insertion} transforms as $\delta_\epsilon V''(p_i) = 2\epsilon  V''(p_i)$, thanks to the implicit $\sqrt{g}$ included in the integral over $x$ and the two fields included when we use the $\phi^4$ vertex \eqref{phi4}. Note that the same $\sqrt{g}$ implies that $\delta(p_1+\cdots+p_n)$ transforms with a factor of $4\epsilon$. Therefore we see that  integrals over momentum must transform as
\be 
\label{intq}
\delta_\epsilon\! \int_q = -4\epsilon\!  \int_q\,,
\ee 
to be consistent with $\delta(q)$. 
Putting all this together we see that $A(p_1,\cdots,p_n)$ transforms with a $(2n-4)\epsilon$ factor, and thus the well-defined vertices $\mathcal{V}_n[\phi]$ transform with a $(-4n+2n+4+2n-4)=0$ factor 
(where the contributions from \eqref{vertex} are listed in order), 
\ie are indeed invariant.

This is not true however of the $n\le2$ contributions. 
We do not consider the case of $\mathcal{V}_0$, 
which only yields a field-independent contribution.
For the vertex $\mathcal{V}_{1}[\phi]$, the Feynman integral is quadratically divergent:
\be 
A = \int_q \frac{1}{q^2}\,.
\label{quadratic}
\ee
If we use a scale-free regularisation such as 
dimensional regularisation, then by dimensions the only possible answer is $A=0$. For a physical regulator such as a UV momentum cutoff $q\le\Lambda$, 
the result is a $\Lambda^2$ divergence that we have to remove by a local counterterm.
It may seem that by putting such a mass counterterm in the bare action we are actually defeating our purpose of starting with a scale-invariant classical theory. However, one must recall that the counterterm, like the loop correction, is of order $\hbar$, so that in the classical limit the bare action is indeed scale invariant.
Since the counterterm is arbitrary, the renormalized mass is also arbitrary. As discussed in the introduction, there is a special theory that preserves scale invariance as much as possible. 
This corresponds to choosing the counterterm to be equal and opposite to the integral (\ref{quadratic}), so that scale invariance is restored for $\mathcal{V}_{1}[\phi]$, in this case by setting it to zero.

Let us now come to the case $n=2$. The integral over $q$:
\be 
\label{beta}
A(p_1,-p_1) = \int_q \frac{1}{q^2 (q+p_1)^2}\,,
\ee
is ultraviolet divergent and thus not well defined. 

As above, we will simply cut off the integral at $|q|=\Lambda$ for large $\Lambda$. 
Now the action of $\delta_\epsilon$ on \eqref{beta} picks up the boundary contribution
\be
\label{boundary}
( +\epsilon\Lambda) \frac{2\Lambda^3}{(4\pi)^2} \frac{1}{\Lambda^4}\ .
\ee
To see this, note first that formally the $\epsilon$ contributions cancel, in the same way that they did rigorously for the $n>2$ cases. The sole contribution thus comes from the boundary. 
Write the $q$ integral as an integral over angles and over the radial direction $|q|$. 
By \eqref{q2}
we are instructed to replace $|q|$ with $(1-\epsilon) |q|$ wherever we see it. But that implies that the ultraviolet boundary to the integral is now at $(1-\epsilon) |q| = \Lambda$, or what is the same: $ |q| = \Lambda (1+\epsilon)$.
The first factor in \eqref{boundary} is this extra contribution from the boundary, and the other factors are from the integrand as a function of $|q|$, in particular
the second factor is the volume of the 3-sphere at $|q|=\Lambda$ divided by $(2\pi)^4$, and the final factor is the contribution from the integrand at the boundary (where since $|p_1|\ll \Lambda$ we can neglect $p_1$). Together with the $\lambda^2/4$ from the two insertions \eqref{insertion}, and the $-1/4$ from \eqref{vertex},
this gives 
\be
\delta_\epsilon\Gamma
= -\epsilon\int_x \frac{\lambda^2}{128\pi^2}\,\phi^4\ ,
\label{des}
\ee
which agrees with (\ref{deltaGamma1},\ref{deltaGamma2}), once we recall that to one loop, the $\beta$ function is:
\be
\label{betaLambda}
\beta=\frac{3\lambda^2}{16\pi^2}\,.
\ee

Just as with $\mathcal{V}_1$, since \eqref{beta} is UV divergent, we need to modify the bare action. 
Adding to it the  counterterm
\be 
\label{counter4}
+\int_x \frac{\lambda^2}{128\pi^2}\, \log\!\left(\frac{\Lambda}{\mu}\right) \phi^4\ ,
\ee
where $\mu$ is the usual arbitrary finite reference scale, ensures that overall the result  is finite. Note however that under the global 
scale transformation \eqref{scaletr}, the counterterm is invariant. 
Thus the renormalized contribution still breaks scale invariance with the same result, namely \eqref{des}. 

On the other hand, now that the total contribution to the EA 
is finite, the breaking can be understood in a different way. The scale $\Lambda$ has disappeared, but scale invariance is still broken, because of the appearance of the scale $\mu$. 
To see that \eqref{breaking} emerges again,
we note that by dimensions the amplitude \eqref{beta} is proportional to $\log(p_1^2/\mu)$. In fact,
the finite part of $\mathcal{V}_2$ is 
\be 
\label{nonlocalphi4}
+\int_x \frac{\lambda^2}{256\pi^2}\, \phi^2 \log\!\left(\frac{-\partial^2}{\mu^2}\right) \phi^2\,,
\ee
up to a local, scale invariant $\phi^4$ term which can be absorbed by the renormalization scheme. By \eqref{q2},  $\delta_\epsilon$ clearly gives again minus the $\beta$-function times the $\phi^4$ operator.
In contrast to the quadratic mass term, this cannot be removed by a local counterterm.

Note that the $\beta$-function is arising in a different way from the RG treatment.  
In the RG treatment, we associate the $\beta$-function as arising not directly from the integral \eqref{beta} but from the counterterm \eqref{counter4} required to make it finite. Indeed the $\beta$-function for the renormalised coupling $\lambda(\mu)$ arises from the requirement that the bare coupling $\lambda(\Lambda)$ is independent of $\mu$, where the bare coupling is the coupling in $S(\phi)$, and from \eqref{counter4} is now given by:
\footnote{\label{footnote:dimensionless} Strictly speaking the notation $\lambda(\mu)$ cannot be correct: a dimensionless variable cannot depend on a dimensionful variable only. It must also depend on a second dimensionful variable
and then through a dimensionless combination of the two.
Thus $\lambda(\mu/\Lambda_\text{dyn})$ (where $\Lambda_\text{dyn}$ is some dynamical scale) would be a better notation.
We stick here to the notation that is common in the QFT literature.}
\be 
\lambda(\Lambda) = \lambda(\mu) + \frac{3\lambda^2(\mu)}{(4\pi)^2} \log\!\left(\frac{\Lambda}{\mu}\right)\,.
\label{Sophie}
\ee

We see in \eqref{nonlocalphi4} that $\mathcal{V}_{2}[\phi]$ is non-local. 
Note that the $\mathcal{V}_{n>2}[\phi]$ are also non-local, as they must be by dimensions. For $\lambda\phi^4$ theory, the $n\!>\!2$ terms contain $2n$ fields and thus the vertex is a negative dimension function of the momenta $p_i$ and clearly therefore must be non-local. 
\footnote{Alternatively we can see this by noting that for all the $n\!\ge\!2$ vertices, there is no Taylor expansion in the external momenta. The integrals from \eqref{A} that would give the coefficients of such a Taylor expansion are all infrared divergent and thus do not exist.} 
We will see in sec. \ref{sec:derivative} why these observations are important for the trace anomaly.

\begin{figure}[ht]
\centering
\includegraphics[scale=0.35]{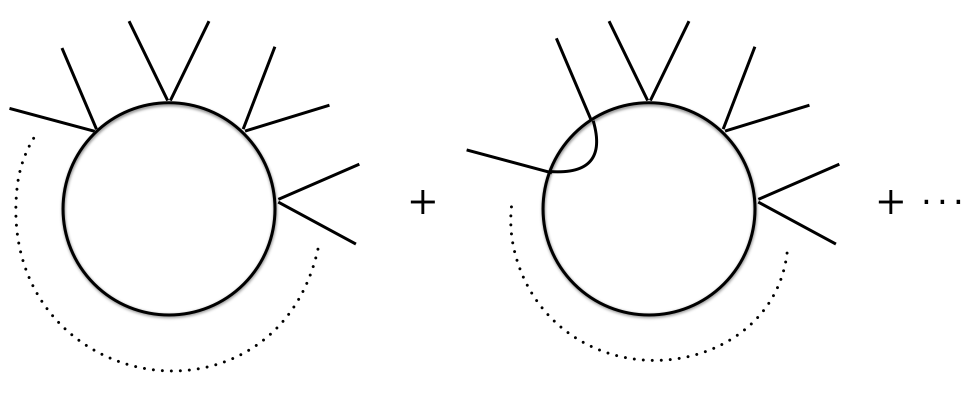}
\caption{How higher-point vertices contribute to the trace anomaly beyond one loop.}
\label{fig:tildelambdarunning}
\end{figure}

Finally note that according to \eqref{deltaGamma2}, the 
$\mathcal{V}_{n>2}[\phi]$ vertices should also contribute to the anomaly, since they are proportional to $\lambda^{n}$. This is true, however since $\beta$ already contains $\hbar$ (starts at one loop), this is a higher loop effect that is thus neglected in this one loop computation (whereas \eqref{des} is a one loop effect on top of the classical contribution 
\eqref{phi4}). Indeed these contributions begin to show up once we include the diagrams shown in fig. \ref{fig:tildelambdarunning}.

\section{The WI of global scale transformations}

\subsection{Derivation}

We now come to the Effective Average Action (EAA),
which is defined as follows.
Let
\bea
W_k(j;g_{\mu\nu})
&=&\log
\int (d\phi)\,
\exp\!\Big[-S
-S_k
+\int \!\! d^dx\,\, j\cdot \phi
\Big]
\eea
be the generating functional obtained
from the action $S+S_k$, where
\bea
S_k(\phi;g_{\mu\nu})&=&
\frac{1}{2}\int_x
\phi R_k(\Delta) \phi
\label{gencutoff}
\eea
and $R_k(\Delta)\equiv k^2 r(y)$,
with $y=\Delta/k^2$ and $\Delta=-\partial^2$,
is a kernel suppressing the contribution
of modes with momenta lower than $k$.
It is quadratic in the fields and only affects the
propagator.
 
The EAA is defined as a modified Legendre transform
\be
\Gamma_k(\phi;g_{\mu\nu})
=-W_k(j;g_{\mu\nu})
+\int_x j\, \phi
-S_k(\phi;g_{\mu\nu})
\label{eaa}
\ee
where $\phi$ denotes here, 
by a slight abuse of language,
the classical VEV of the corresponding quantum field;
the sources have to be interpreted
as functionals of these classical fields
and the last term subtracts the cutoff that had been
inserted in the beginning in the bare action.

The main virtue of this functional is that it satisfies
a simple equation \cite{Wetterich:1992yh,Morris:1993qb,Morris:1994ki}
\be
\label{erge}
k\frac{\partial\Gamma_k}{\partial k}
=\frac{1}{2}\Tr\left(\Gamma_k^{(2)}+R_k\right)^{-1}
k\frac{\partial R_k}{\partial k}
\ee
where $\Gamma_k^{(2)}$ is the second variation of the EAA
with respect to the field.
We note that this equation knows nothing about
the action that entered in the functional integral.
\footnote{In principle the bare action could be reconstructed
from the limit of \emph{a given solution} $\Gamma_k$ for $k\to\infty$ \cite{Morris:1993qb,Manrique:2008zw,Morris:2015oca}.}
In particular, if we assume that $S$ is scale invariant, 
as we shall henceforth do, (\ref{erge})
remains exactly the same.

We can now calculate the transformation of the cutoff term
under rescaling.
The Laplacian contains an inverse metric and therefore
transforms under (\ref{scaletr}) by
\be
\delta_\epsilon\Delta=-2\epsilon\Delta\ .
\label{laptrans}
\ee
Since $k$ does not change,
we find $\delta_\epsilon R_k=-2\epsilon k^2 y r'$.
On the other hand $\partial_t R_k=2 k^2 r-2k^2 y r'$,
so
\be
\delta_\epsilon R_k=\epsilon(-2R_k+\partial_t R_k)\ .
\label{piero}
\ee
When we apply the variation to the cutoff action,
all terms cancel except for the last term in (\ref{piero}),
giving
\be
\delta_\epsilon S_k=
\frac{\epsilon}{2}\int_x
\phi\, \partial_t R_k \phi\ ,
\ee

We now have all the ingredients that are needed to derive the WI.
We subject $W_k$ to a background scale transformation,
with fixed sources and fixed $k$.
Since the action $S$ is assumed invariant,
the only variations come from 
the measure, the cutoff and source terms:
\bea
\delta_\epsilon W_k&=&
\cA(\epsilon)
+\langle\delta_\epsilon S_k\rangle
+\int_x j \langle\delta_\epsilon\phi\rangle
\nonumber\\
&=&
\cA(\epsilon)
+\epsilon\left[
-\frac{1}{2}\int_x
\frac{\delta W_k}{\delta j}\partial_t R_k
\frac{\delta W_k}{\delta j}
+\frac{1}{2}
\Tr\,\partial_t R_k
\frac{\delta^2 W_k}{\delta j\delta j}
+\int_x j d_\phi\frac{\delta W_k}{\delta j}\right]\ .
\label{varW}
\eea
Here the first term comes from the variation of the measure
and coincides with the trace anomaly that one always finds in the EA.
It can be calculated for example by Fujikawa's method.
\footnote{This is a somewhat abstract interpretation
that stands for whatever UV regularization one is using.
For example, it can be calculated by using
an UV momentum cutoff, as we saw in section 2.}
The second term comes from the variations of the
cutoff and the last comes from the variation of the source terms.
The variation of the EAA can be computed from (\ref{eaa}):
\be
\delta_\epsilon\Gamma_k=
-\delta_\epsilon W_k
+\int_x j \delta_\epsilon\langle\phi\rangle
-\delta_\epsilon S_k(\langle\phi\rangle)\ .
\label{varGamma}
\ee
Using (\ref{varW}), the source terms cancel out
(since the variation is linear in the field
we have 
$\langle\delta_\epsilon\phi\rangle
=\delta_\epsilon\langle\phi\rangle$)
and the first term in bracket in the r.h.s. of (\ref{varW}) cancels out
with the last term in (\ref{varGamma}).
The middle term in the same bracket can be rewritten in terms of $\Gamma_k$
yielding
\bea
\delta_\epsilon \Gamma_k&=&
-\cA(\epsilon)
+\epsilon\,\frac{1}{2}
\Tr\left(
\frac{\delta^2\Gamma_k}{\delta\phi\delta\phi}
+R_k\right)^{-1}\partial_t R_k
\label{wir}
\eea
Apart from the factor $\epsilon$, the second term is exactly the
FRGE.
We can thus write as in \eqref{wi}:
\be
\nonumber
\delta_\epsilon \Gamma_k=-\cA(\epsilon)
+\epsilon\partial_t\Gamma_k\ .
\ee
This is our main result.
We see that in addition to the standard trace anomaly,
which originates from the UV regularization,
there is another source of variation due to the IR cutoff $k$, 
that is exactly proportional to the FRGE.

\subsection{Applications}

Let us recall how the WI (\ref{deltaGamma1},\ref{deltaGamma2}) 
is often used in practice.
If $\beta(\lambda)$ is known and if there is a sole reason for the breaking of scale invariance \eg through a vacuum expectation value $\langle\phi\rangle$, or through constant background scalar curvature $\cR$, or through summarising external momentum dependence 
by $-\partial^2$, then (\ref{deltaGamma1},\ref{deltaGamma2}) can be integrated to give the exact answer for the physical EA $\Gamma$ in terms of these quantities, provided its dependence on $\lambda$ is also already known.\footnote{To be clear, for the $\cR$-dependence we are again considering only the interacting part of the anomaly \cite{Drummond:1977dg}.} 

To see this, let $\chi$ be the sole reason for breaking of scale invariance. Without loss of generality, we can set $\delta_\epsilon \chi = -\epsilon\chi$. Thus in the above examples we have chosen $\chi$ to be  $\langle\phi\rangle$ (in $d=4$ dimensions), or $\sqrt{\cR}$, or $\sqrt{-\partial^2}$ respectively. Then
\be 
\label{deltaChi}
\delta_\epsilon \Gamma = -\epsilon\, \chi  \partial_\chi \Gamma\,.
\ee
Combining this equation with (\ref{deltaGamma1},\ref{deltaGamma2}) tells us that 
\be 
\label{GammaChi}
\Gamma = \Gamma\left(\lambda(\chi)\right)\,, 
\ee
\ie depends on $\chi$ only through its dependence on $\lambda$, where we suppress the dependence of $\Gamma$ on the other quantities.
In general $\lambda(\chi)$ is given by the implicit solution 
of the RG equation:
\be
\label{implicit}
  \int^{\lambda(\chi)}_{\lambda(\mu)} 
  \frac{d\lambda}{\beta(\lambda)} = \log(\chi/\mu)\,.
\ee
In a one-loop approximation this can be solved explicitly:
\be 
\label{lambdaChi}
\lambda(\chi) = \lambda(\mu) +\frac{3\lambda^2(\mu)}{(4\pi)^2}\log\left(\frac{\chi}{\mu}\right)\,.
\ee 
Evidently \eqref{GammaChi} guarantees the standard form for the trace anomaly, since operating with  \eqref{deltaChi} takes us back to \eqref{deltaGamma2}. 

As a concrete example, equations \eqref{GammaChi} and \eqref{lambdaChi}
imply that by setting $\chi=\langle\phi\rangle$
in the tree-level term \eqref{phi4}, one obtains the Coleman-Weinberg potential:
\be 
\label{CW}
V(\langle\phi\rangle) = \frac{\langle\phi\rangle^4}{4!}\left( \lambda(\mu) 
+\frac{3\lambda^2(\mu)}{(4\pi)^2}\log(\langle\phi\rangle/\mu)\right)\,.
\ee
Similarly by setting $\chi^2=\cR$ one gets
the interacting part of the conformal anomaly 
on a spherical background \cite{Drummond:1977dg}.

Let us now come to the EAA.
If we again assume that there is 
just one source of breaking of scale invariance,
we have also in this case
\be 
\label{deltaChik}
\delta_\epsilon \Gamma_k = -\epsilon\, \chi  \partial_\chi \Gamma_k\,,
\ee
Then combining this with the WI \eqref{wi}
and with \eqref{deltaGamma2},
that we will show at the end of sec. 4.2 to hold also when $k>0$,  we have
\be 
\left[\beta(\lambda)\partial_\lambda - \chi\partial_\chi - \partial_t \right]\Gamma_k = 0\,.
\ee
This equation can be solved, \eg by the method of characteristics. 
The solution implies that $\Gamma_k$ also has a restricted functional form, which can for example be written as:
\be 
\label{GammaTwoArgs}
\Gamma_k = \hat{\Gamma}\left(\vphantom{\hat{\Gamma}}{\chi}/k\,,\,\lambda\!\left(\chi^ak^{1-a}\right)\right)\,,
\ee
for any number $a$.
We shall see in Section 5 how this reduces,
in the limit $k\to0$, to \eqref{GammaChi}.

Notice that the anomalous WI \eqref{wi} does not
give us any information on the dependence of $\hat\Gamma$
on its arguments, nor on the dependence of $\lambda$ on $k$.
This information has to be obtained by other means,
{\it e.g.} by solving the FRG in some approximation.

\section{Approximations}

\subsection{One-loop EAA}
\label{sec:EAA}

We shall now see how the WI looks like at one loop and how the anomaly is recovered in this approximation.
In the spirit of the FRGE, we could compute first 
$\partial_t\Gamma_k$ and then integrate from some UV scale 
$\Lambda$ down to $k$ to obtain $\Gamma_k$. 
In practice, in the one loop approximation, this is equivalent to
just repeating the calculation of Section 2 with the IR regulator
in the bare action.

The two-point function is independent of the external momentum
and is a simple mass term. The quantum correction depends on the form of the regulator but is generally of the form $B\Lambda^2$, where $B$ is a scheme-dependent coefficient, plus other (possibly logarithmically divergent) terms depending also on $k$. For example if we use the optimized cutoff $R_k(z)=(k^2-z)\theta(k^2-z)$ we have
\be
m^2(k)=m^2(\Lambda)+\frac{\lambda}{32\pi^2}\left[\Lambda^2
-\frac{1}{2}k^2\right]\ .
\ee
The initial value for the mass at the UV scale now plays the role of the counterterm. One can choose it so that the renormalized mass
$m_R^2=m^2(k=0)$ has any value.
In particular we can fine-tune it so that
the renormalized mass is exactly zero. This defines the critical trajectory.
It is important to note that even though this choice eliminates this source of scale symmetry breaking at $k=0$, for $k\not=0$ we have
\be
m^2(k)=-\frac{\lambda}{64\pi^2}k^2\ .
\ee
This is then the main source of scale symmetry breaking for $k\not=0$,
being of order $\lambda$,
while the anomaly (\ref{breaking}) is of order $\lambda^2$,
as is seen from (\ref{betaLambda}).

Let us now come to the four point function, that contains 
operators of the form
$\phi^2(-\partial^2)^{n}\phi^2$.
At one loop these all arise from inserting the IR regularisation into \eqref{beta} to give:
\be 
A_k(p,-p) = \int_q \frac{1}{\left[q^2+R_k(q)\right]\left[ (q+p)^2+R_k(q+p)\right]}\,,
\ee
In general this integral is quite complicated,  
but since it is only logarithmically divergent we can get away with choosing the simple momentum independent $R_k = k^2$, \ie a $k$-dependent mass term.  This IR cutoff is not strong enough to work with more severe cases, but by inspecting this example we will be able easily to see what the general $R_k$ will give. Using the Feynman trick the integral is
\be 
A_k(p,-p) = \int_0^1\!\!d\alpha\int_q \frac{1}{\left[ (1-\alpha) q^2 + \alpha (q+p)^2+k^2\right]^2}\,.
\ee 
Completing the square and shifting internal momentum we get
\be 
A_k(p,-p) = \int_0^1\!\!d\alpha\int_q \frac{1}{\left[ q^2 +k^2 + (1-\alpha)\alpha\, p^2\right]^2}\,.
\ee
This integral is now subject to the UV boundary condition that $|q-\alpha p| \le \Lambda$, but replacing it with $|q|\le \Lambda$ only introduces an error of order $p^2/\Lambda^2$ which vanishes as we take the UV limit. 
Performing the $q$ integral we thus find that the EAA contains
\be 
\int_x
\frac1{8\pi^2}\left\{ \log\left(\frac{\Lambda^2}{k^2}\right) -1- \int_0^1\!\!\!d\alpha\, \log\left[1+ (1-\alpha)\alpha\, p^2/k^2\right] \right\}\,,
\ee
where again we discard terms that vanish as $\Lambda\to\infty$. Recalling from above \eqref{des}, the factor of $-\lambda^2/16$, we see that the same counterterm \eqref{counter4} 
will render this finite. 
Indeed including the $\phi^4$ contribution, \eqref{Sophie}, from the bare action $S[\phi]$, 
we have altogether:
\be 
\label{phi4ops}
\int_x\left\{
\left[\lambda(\mu) + \frac{3\lambda^2(\mu)}{32\pi^2}
\left(1+\log\left(\frac{k^2}{\mu^2}\right)\right)\right]\frac{\phi^4}{4!}+\frac{\lambda^2(\mu)}{256\pi^2}\, \phi^2\! \int_0^1\!\!\!d\alpha\, \log\left[1- (1-\alpha)\alpha\, \partial^2/k^2\right] \phi^2
\right\}\,
\ee
however for the FRG, it is more natural to choose as renormalization condition that the coefficient of $\phi^4/4!$ is the renormalized coupling $\lambda(k)$. 
In this way it is clear what it means to pick a solution that breaks scale invariance the least: we should pick the solution that breaks the invariance only through the running of this coupling \cite{Morris:1998da}. It implies that the bare coupling $\lambda(\Lambda)$ is now set equal to:
\be 
\label{bareVsRenk}
\lambda(\Lambda) = \lambda(k) + \frac{3\lambda^2(k)}{32\pi^2} \left\{\log\!\left(\frac{\Lambda^2}{k^2}\right)-1\right\}\,.
\ee
In this way we avoid introducing an explicit extra scale $\mu$, whilst \eqref{bareVsRenk} and the $\beta$-function \eqref{betaLambda}, now tells us that $\lambda(\Lambda)$ is independent of $k$ up to terms of higher order, as it should be. Then \eqref{phi4ops} just reads:
\be
\int_x\left\{
\frac{\lambda(k)}{4!}\,\phi^4+\frac{\lambda^2(k)}{256\pi^2}\, \phi^2\! \int_0^1\!\!\!d\alpha\, \log\left[1- (1-\alpha)\alpha\, \partial^2/k^2\right] \phi^2\right\}\,,
\ee
after using $\lambda^2(\mu)=\lambda^2(k) + O(\lambda^3)$. Finally, Taylor expanding the last term  gives us the derivative expansion we were aiming for:
\be 
\label{derivphi4}
\int_x\left\{
\frac{\lambda(k)}{4!}\,\phi^4+ \frac{\lambda^2(k)}{256\pi^2} \sum_{n=1}^\infty a_n\ \phi^2 \left(\frac{-\partial^2}{k^2}\right)^{\!n}\!\!\phi^2\right\}\,,
\ee
where we learn that with $R_k = k^2$, the $a_n$ are given by:
\be 
\label{an}
a_n = (-1)^{n+1} \,\frac{n!(n-1)!}{(2n+1)!}\,.
\ee
Let us rewrite \eqref{wi} in the form
\be 
\label{anomaly}
\cA(\epsilon)=
\epsilon\partial_t\Gamma_k-\delta_\epsilon \Gamma_k\,.
\ee
Acting with $\epsilon\partial_t-\delta_\epsilon$ 
on the second term in (\ref{derivphi4}) gives zero.
Acting on the first term, $\delta_\epsilon$ gives zero and
$\epsilon\partial_t$ reproduces exactly (\ref{breaking}).
We thus see how the one-loop EAA (\ref{derivphi4})
computed with a simple mass-like cutoff reproduces the anomaly.
For other forms of the cutoff function $R_k$ the coefficients $a_n$ would have a different form, but the calculation would proceed exactly in the same way, leading always to the same final form 
\eqref{breaking} for the anomaly.

The one loop EAA also has  $2n$-point vertices   
$\mathcal{V}_{k,\,n}[\phi]$ where $n\ne2$. Their expansion in local operators gives powers of derivatives and the field balanced by powers of $k$ according to dimensions. (For $n=1$ the tadpole integral yields exclusively a mass term proportional to $k^2$.) As at the end of sec. \ref{sec:Feynman}, at one loop the $\lambda^n$ factor does not run, being already proportional to $\hbar$. Therefore, as we will see confirmed also in the next section, the application of the right hand side of \eqref{anomaly} just gives zero. From the (derivative expansion of the) whole of the one loop EAA, we are therefore left just with the one contribution coming from \eqref{breaking}, which here is reproduced entirely from the RG running of the $\lambda(k)\phi^4$ term.

\subsection{Local expansions}
\label{sec:derivative}

In practical applications of the EAA, one often assumes that
it has the form
\be
\Gamma_k=\sum_i\lambda_i(k)\,\cO_i\ ,
\label{sl}
\ee
where the $\cO_i$ are integrals of local operators constructed with the
fields, the metric and derivatives.
For the purpose of counting, notice that the integral
contains $\sqrt{g}$ and therefore carries $d/2$ powers of the metric.
Generically such approximations are called ``truncations''.
Systematic expansions are the derivative expansion
and the vertex expansion,
in which cases the sum in (\ref{sl}) is infinite
and contains arbitrary powers of the field
or of the derivative, respectively \cite{Morris:1994ie}.

Differently from the previous section, we are here treating each operator as having its own separate coupling $\lambda_i$, and
absorbing all powers of $k$ into these couplings. 
Later on, we will specialise to the case where the continuum limit is controlled by just one marginal coupling.

For the WI \eqref{wi}, it is enough to consider one monomial at the time.
Let $\cO_i$ involve $n_\phi$ powers
of $\phi$ and, in total, $n_g$ powers of the metric. 
The scaling dimension of $\cO_i$ under \eqref{scaletr}, is  
\be
\Delta=-2n_g+\frac{d-2}{2}n_\phi
\label{delta}
\ee
and the scaling dimension of $\lambda_i$ under \eqref{scaletr} (which is minus its mass dimension) is $-\Delta$.
We can also write
$$
\lambda_i\,\cO_i=\tilde\lambda_i\,\tilde\cO_i
$$
where
\be
\tilde\lambda_i=k^\Delta\lambda_i\ ;\qquad
\tilde\cO_i=k^{-\Delta}\cO_i\,,
\label{tilded}
\ee
which thus implies that $\tilde\lambda_i$ is dimensionless.
The l.h.s. of the WI is
\be
\label{deltaDerivExpansion}
\delta_\epsilon\left(\lambda_i\,\cO_i\right)=
\epsilon\lambda_i\left(2n_g-\frac{d-2}{2}n_\phi\right)\cO_i
=-\epsilon\,\Delta\,\lambda_i\,\cO_i
\ee
On the other hand, one has
\be
\partial_t\left(\lambda_i\,\cO_i\right)
=\partial_t\lambda_i\,\cO_i\ ,
\ee
since the dimensionful operator in itself has no dependence on $k$.
Thus the WI gives
\be
-\epsilon\lambda_i\Delta\,\cO_i=
-\cA(\epsilon)
+\epsilon\partial_t\lambda_i\,\cO_i
\ee
Bringing the l.h.s. to the r.h.s. it reconstructs the
derivative of $\tilde\lambda_i$, times $k^{-\Delta}$,
which can be rewritten as
\be
\cA(\epsilon)=\epsilon\,\partial_t\tilde\lambda_i\,\tilde\cO_i\ .
\ee
Thus for an action of the form (\ref{sl})
the WI implies
\be
\label{betasum}
\cA(\epsilon)=\epsilon\,\sum_i\tilde\beta_i\,\tilde\cO_i
\ .
\ee
We see from this expression that the anomaly 
receives contributions from all the operators.
Let us also note here that if the effective action
has an (infinite) expansion of the form \eqref{sl},
and if one keeps {\sl all} the terms, this formula is not
an approximation anymore.

Now let us consider what form (\ref{betasum}) must take when the continuum limit is controlled by a single marginal coupling $\lambda$, as in the one loop calculations in the previous sections. In the expansion over the local operators $\cO_i$ we have identified couplings $\lambda_i(k)$ as the parameters conjugate to these local operators. One of these $\lambda_i$ is the coupling $\lambda$ itself. In the critical continuum limit the other couplings 
are not independent but are functions of $\lambda$ and $k$. The 
scaled couplings $\tilde{\lambda}_i=\tilde{\lambda}_i(\lambda)$ are dimensionless and thus have no explicit $k$ dependence. They gain their $k$ dependence only through their dependence on $\lambda$. (Note that, being marginal, $\tilde{\lambda}= \lambda$.) Therefore
\be 
\label{tbi}
\tilde{\beta}_i = \partial_t\tilde{\lambda}_i = \partial_t\lambda \partial_\lambda\tilde{\lambda}_i = \beta(\lambda)\, \partial_\lambda\tilde{\lambda}_i\,.
\ee 
Notice that \eqref{tbi} also holds for the coupling $\lambda_i$ that is $\lambda$ itself since in this case it simply says $\tilde{\beta}_i = \beta_i = \beta$. Then \eqref{betasum} can be rewritten as 
\be 
\cA(\epsilon) =\epsilon\, \beta(\lambda)\, \partial_\lambda \sum_i \tilde{\lambda}_i\,\tilde{\cO}_i\,,
\ee
or simply
\be 
\label{AinGamma}
\cA(\epsilon) =\epsilon\, \beta(\lambda)\, \partial_\lambda \Gamma_k\,.
\ee
We see that this is \eqref{deltaGamma2} that applies to the EA, and moreover it holds also at finite $k$ \ie for the EAA. What does not hold at finite $k$, is \eqref{deltaGamma1}, namely the statement that $\delta_\epsilon$ induces the anomaly 
$\cA(\epsilon)$ and only this anomaly.

\section{Recovering the standard form of the trace anomaly}

To recover \eqref{deltaGamma1}, we need to study the limit $k\to0$, keeping all other quantities fixed. In this limit $\Gamma_k\to\Gamma$, and the breaking due the IR cutoff $R_k$ should disappear. Comparing \eqref{wi} and \eqref{deltaGamma1},  we see that this is true provided that $\partial_t \Gamma_k\to0$, which indeed must also hold in this limit as we show below. Note that the derivative expansion, or any approximation of $\Gamma_k$ in terms of local operators, implies a Taylor expansion of the vertices in $p_i/k$, the dimensionless momenta. Since we hold $p_i$ fixed and let $k\to0$, such approximations are not valid in the regime we now need to study.

We will get insight by first inspecting the one loop case. At one loop, the term that contains the trace anomaly in the small $k$ limit is in fact the non-local term \eqref{nonlocalphi4}, which indeed is missing from any local approximation to the EAA, 
in particular from the derivative expansion considered in the last section. 
Together we therefore have:
\be 
\label{nonlocalphi4k}
\Gamma_k\ \ni\  \frac1{4!}\int_x\left\{ \lambda(k)\, \phi^4\,+\, \frac{3\lambda^2(k)}{32\pi^2}\, \phi^2 \log\!\left(\frac{-\partial^2}{k^2}\right) \phi^2\right\}\,,
\ee
where the explicit $k^2$ is supplied by the counterterm in \eqref{bareVsRenk}, in preference to the $\mu^2$ supplied by \eqref{Sophie}. 
Now note that with the non-local term included, the $\phi^4$ coefficient is actually independent of $k$ (to the one-loop order in which we are working), the $\beta$-function contribution cancelling against the explicit $k$ dependence in \eqref{nonlocalphi4k}. We see that the non-local term is just what is needed in order to ensure that $\Gamma_k$ has a sensible limit. Indeed we could have found the non-local term by insisting that $\Gamma_k$ becomes independent of $k$ as $k\to0$, holding everything else finite. 
This implies a practical method for recovering such non-local terms from the flow of the couplings,  
as we will see shortly. 
By adding the missing non-local term as in \eqref{nonlocalphi4k}, we now have a four-point vertex that satisfies 
\be 
\partial_t \Gamma_k^{(4)} = 0\,,
\ee
but also gives the standard form of the trace anomaly. In this way we have reproduced \eqref{deltaGamma1}.

Let us now set ourselves in the situation when there is
a single source of scale symmetry breaking $\chi$, as in section 3.2.
To get an explicit answer for the four-point vertex in the limit as $k\to0$, we can for example solve for $\lambda(k)$ in terms of some $\lambda(\mu)$. The solution is just \eqref{lambdaChi} with $\chi$ replaced by $k$. Plugging this back in \eqref{nonlocalphi4k} we get the same expression as \eqref{nonlocalphi4k}, but with 
$-\partial^2$ replaced by $\chi^2$ and $k$ replaced by $\mu$, a consequence of the fact that  the physical EA is  actually an RG invariant, and thus independent of $k$ or $\mu$.

Now, if we want to go beyond the one loop approximation,
in general we will have to solve equation \eqref{implicit}.
However in perturbation theory, by iteration, we can explicitly find this form of the solution. For example to two loops, writing 
\be 
\label{betag}
 \beta = \beta_1 \lambda^2 + \beta_2 \lambda^3\,,
\ee
where $\beta_1$ is the coefficient in \eqref{betaLambda},
we must have
\be 
\label{lambdaTwoloops}
\lambda(\chi) = \lambda(k) +\beta_1 \lambda^2(k) \log( \chi/k)\, + \lambda^3(k)\, \alpha(\chi/k)\,,
\ee
for some function $\alpha$ to be determined, where we recognise that \eqref{nonlocalphi4k} and \eqref{betaLambda}  already fix the $\beta_1$ term. Differentiating the above with respect to $t$, using \eqref{betag}, and requiring that overall the result vanishes, we confirm again the $\lambda^2$ piece, and determine that
\be 
\partial_t \alpha(\chi/k) = -\beta_2 -2\beta^2_1\log(\chi/k)\,.
\ee
Integrating we thus have
\be
\label{a} 
\alpha(\chi/k) =  \beta_2 \log( \chi/k) +\beta_1^2 \log^2( \chi/k)\,.
\ee
Note that the integration constant vanishes since by \eqref{lambdaTwoloops}, we must have $\alpha(1)=0$.
Clearly our EA, \eqref{GammaChi}, then does satisfy the anomalous WI, provided we recall that in \eqref{deltaGamma2} we have $\lambda=\lambda(\chi)$. In particular this means that the trace anomaly appears at this order as  \eqref{deltaGamma2} where however
\be 
\label{betaExpanded}
\beta = \beta_1  \lambda^2(\chi) + \beta_2 \lambda^3(\chi) = \beta_1\lambda^2(k) +\lambda^3(k) \left\{\beta_2 +2\beta^2_1\log( \chi/k)\right\}\,.
\ee

By design, and despite appearances,  $\lambda(\chi)$ and  $\Gamma$, are independent of $k$. Indeed, from \eqref{lambdaTwoloops} and \eqref{a}, we know that substituting 
\be 
\label{ktomu}
\lambda(k) = \lambda(\mu) +\beta_1 \lambda^2(\mu) \log( k/\mu)\, + \beta_2\, \lambda^3(\mu)\log(k/\mu) + \beta_1^2\, \lambda^3(\mu) \log^2(k/\mu)\,,
\ee
into \eqref{lambdaTwoloops} and \eqref{betaExpanded}, 
will eliminate $k$ and $\lambda(k)$, in favour of $\mu$ and $\lambda(\mu)$, making explicit the fact that these formulae are actually independent of $k$.

Finally, we add a note to clarify the r\^ole of $\mu$. Recall that the running of couplings
with respect to the scale $\mu$,
is fundamentally different from the running of $k$ in the Wilsonian RG. Whereas $k>0$ parametrises an infrared cutoff, meaning that there are still low energy modes to be integrated out, $\mu$ is a dimensional parameter that remains even when the functional integral is completed. Then the RG is realised through $\mu$, however only in the Callan-Symanzik sense: physical quantities must actually be independent of $\mu$.  
To the extent that the EA is a physical quantity, the EA must therefore also be independent of $\mu$. In this sense, dependence on $\mu$ is fake. It, and $\lambda$, can be eliminated in favour of a fixed dynamical scale (completing the so-called dimensional transmutation, \cf footnote \ref{footnote:dimensionless}). In the limit that $k\to0$, we can only be left with this fake dependence on $k$, thus (only) in this limit $k$ and $\mu$ appear on the same footing, as is exemplified explicitly in \eqref{ktomu}.

\section{Fixed points}

There is no connection between scale invariance of the classical
action and scale invariance at the quantum level.
One can have the former without the latter, as we have seen,
but also vice-versa.
Quantum scale invariance is related to the existence of fixed points.
At a fixed point, (\ref{deltaGamma1}) vanishes. 
The scale-invariance of the EA is then explicitly realised as invariance under the transformation $\delta_\epsilon$. Since (\ref{deltaGamma2}) also implies $\beta(\lambda)=0$, $\lambda$ can no longer depend on $\mu$ and becomes a fixed number, as indeed is verified by \eqref{ktomu} since now all the $\beta_n$ vanish. For the same reason, all the explicit $\mu$ dependence also disappears from the EA, as obviously it must in order for the EA to be overall independent of $\mu$. (Again this is verified by \eqref{lambdaTwoloops} and \eqref{a}.)

The connection to scale invariance is less direct for the EAA. In the presence of (generally) dimensionful couplings $\lambda_i$,
a fixed point is defined by the vanishing of the
beta functions of their dimensionless cousins $\tilde\lambda_i$,
as in (\ref{tilded}), {\it i.e.}
$$
\tilde\beta_i(\tilde\lambda_j)=0\ .
$$
One immediate consequence of (\ref{betasum})
is then that the anomaly vanishes at a fixed point.
This however does not lead to the statement that the EAA
is scale-invariant at a fixed point,
according to the definition of scale transformations
that we used so far.
Indeed, if we look at equation (\ref{deltaDerivExpansion})
we see that the variation of the EAA under an infinitesimal
scale transformation $\delta_\epsilon$ is not zero in general.
It is only zero in the case when $\Delta=0$,
{\it i.e.} when all the couplings $\lambda_i$ are 
themselves dimensionless.

Consider, however, a different realization of scale invariance,
namely one where we also transform the cutoff scale by \cite{Morris:2016spn,Percacci:2016arh,Ohta:2017dsq,Nieto:2017ddk}
\be
\label{hatdelta}
\hat\delta_\epsilon k=-\epsilon\,k\ ,
\ee
the action of $\hat\delta_\epsilon$ being the same as
the action of $\delta_\epsilon$ on all other quantities.
Then, instead of (\ref{deltaDerivExpansion}) we have
\bea
\hat\delta_\epsilon\Gamma_k&=&
\delta_\epsilon\Gamma_k
-\epsilon\sum_i k\partial_k\lambda_i \cO_i
\nonumber\\
&=&-\epsilon\sum_i\left(\Delta_i\tilde\lambda_i
+\beta_i k^{\Delta_i}\right) \tilde\cO_i
= -\epsilon\,\sum_i\tilde\beta_i\,\tilde\cO_i\ .
\eea
This implies that at a fixed point one has invariance
under the scale transformations generated by $\hat\delta_\epsilon$.

From this Wilsonian point of view, the relevant notion of
scale transformation is one where the cutoff is also acted upon,
and a fixed point is not a  point where only dimensionless
couplings are present, but rather one where all dimensionful
couplings in the fixed point action appear as (non-universal) numbers times the appropriate power of $k$. It is this fact that ensures that the fixed point action does not vary with $k$, when all variables are written in dimensionless terms (using the appropriate scaling dimensions). Indeed we also recall that when written in these terms, the eigenoperators, which are integrated operators of definite scaling dimension $d_\cO$, correspond to linearised perturbations about the fixed point action whose associated couplings carry power law $k$-dependence, namely $k^{d-d_\cO}$. Thus the behaviour of these linearised couplings under change of scale is entirely given by \eqref{hatdelta}.

We can further clarify the relation to the Wilsonian RG by the following argument. 
The partial derivative $\partial_t$ gives zero when acting
on $\cO$, because all the $k$-dependence is assumed to be
in the coupling, and therefore 
$\partial_t\tilde\cO=-\Delta\tilde\cO$.
Let us make this explicit by writing,
for a monomial $\lambda\cO$ in the EAA:
$$
\partial_t(\lambda\cO) |_\cO
=(\tilde\beta-\Delta\tilde\lambda)\tilde\cO\ .
$$
If instead we take the derivative keeping $\tilde O$ fixed,
we get
$$
\partial_t(\lambda\cO)\big |_{\tilde\cO}
=\tilde\beta\tilde\cO\ .
$$
This implies that for the EAA, which is a sum of terms of this type,
the flow for scaled fields is
\be
\epsilon\partial_t\Gamma_k\big |_{\tilde\cO}
=\epsilon\partial_t\Gamma_k\big |_\cO
-\delta_\epsilon\Gamma_k\ ,
\label{ernesto}
\ee
where (\ref{deltaDerivExpansion}) has been used.
This equation is just the definition of an infinitesimal Wilsonian RG transformation in the way it was originally formulated \cite{Wilson:1973jj}. Thus on the right hand side, the first term is an infinitesimal change of the coarse graining scale
(``Kadanoff blocking''), while the second term is the infinitesimal rescaling back  to the original scale (hence the minus sign).

Now recall that in the WI (\ref{wi}), the $t$-derivative 
is taken at fixed $\cO$:
\be
\cA=\epsilon\partial_t\Gamma_k\big |_\cO
-\delta_\epsilon\Gamma_k \,.
\label{wi2}
\ee
Comparing to \eqref{ernesto}, it is immediate to see  that the anomaly, $\cA$, and the Wilsonian RG transformation, $\epsilon\partial_t\Gamma_k\big |_{\tilde\cO}$, are effectively the same thing.  Indeed this is just eqn. \eqref{betasum} derived in a different way.

\section{Concluding remarks}

One generally speaks of an anomaly when a symmetry of the classical action cannot be maintained in the corresponding quantum theory.
It is implicit that it is desirable to maintain the symmetry as much as possible and a violation of the symmetry can only be accepted when it is unavoidable.
Thus, violations of a symmetry due to a ``bad'' choice of regulator are not usually characterized as anomalies. Likewise when the anomaly can be removed by adding a local counterterm to the action.
In this strict sense, the quadratic renormalization of the mass does not give rise to an anomaly.
In renormalizable field theories in four dimensions, the anomaly is due only to the logarithmic renormalization of a marginal coupling constant.

From the point of view of Wilsonian renormalization,
a theory can be defined as a RG trajectory in the ``theory space'' of all possible effective actions.
Perturbative renormalizability and the symmetries of the bare action are not important.
The only question that is physically relevant
is whether a theory has a symmetry in the quantum sense.
We have shown that the assumption of classical scale invariance
gives rise, in addition to the RG equation, to the WI
(\ref{wi}) that quantifies the amount of scale symmetry breaking.

Given that the RG flow equation is exactly the same whether
or not the classical action is scale invariant,
one may wonder what additional information the WI may have.
To understand this, it is useful to present the WI in the form \eqref{wi2},
showing that the anomaly can be identified with the
Wilsonian definition of RG.
When the theory is classically scale invariant, there is
an independent definition of $\cA$ and \eqref{wi2} is an
identity that can be tested in actual calculations.
When nothing is assumed on the classical action,
one can still assume that \eqref{wi2} holds,
and in this case it becomes the {\it definition} of the
anomaly $\cA$. 
It is in some sense the broadest generalization
of the standard perturbative statement (\ref{breaking}), 
since it applies to any theory, independently
of its UV properties, and it preserves the essential feature
that the anomaly vanishes at a fixed point.
\footnote{For a recent discussion of the scale anomaly
for theories that are not classically scale invariant see
\cite{Casarin:2018odz}.}

One can then distinguish three classes of trajectories.
The mass parameter runs quadratically and generically
ends up with a nonzero value in the extreme infrared.
In these ``gapped'' theories scale invariance is badly broken.
Then there is a subclass of ``critical'' trajectories for which the renormalized mass in the extreme infrared ends up being zero.
These trajectories ``almost'' realize scale symmetry,
but have an anomaly in the strict sense discussed above.
As we have discussed in subsection 3.2, the WI is really only useful when one restricts oneself to such trajectories.
Finally there are trajectories that remain exactly at a fixed point for all scales. These trajectories fully realize quantum scale invariance. 
In the scalar theory in four dimensions that we have considered here as an example, the only such fixed point is the free theory,
but there exist nontrivial fixed points in less than four dimensions and there are other examples of four-dimensional field theories
with nontrivial fixed points \cite{LISA}.

There are relations of this work to several other strands of research and 
various natural extensions.
One extension is to consider the WI of special conformal transformations.
This has been discussed in \cite{Delamotte:2015aaa,Rosten:2014oja,Rosten:2016zap}
and, more specifically in relation to the trace anomaly, 
in \cite{Rosten:2018cyr}.
Another generalization is to make the scale transformations
position-dependent. This can be used as a technical device in flat space physics \cite{Osborn:1991gm} but is most natural
in a gravitational context 
\cite{Codello:2013iqa,Codello:2014wfa}.

Another point to be kept in mind is that interpreting the
renormalization scale as a VEV of a dynamical field
leads to a (typically non-renormalizable) 
theory where scale symmetry is not broken.
This has been discussed recently in \cite{Lalak:2018bow}.
Related observations have been made for local Weyl transformations
in the presence of a dilaton in
\cite{englert,fv,shapo}
and for the EAA in \cite{Percacci:2011uf,Codello:2012sn}.

In a gravitational context, 
the result (\ref{wi}) bears some resemblance to our
earlier results for the WI of split Weyl transformations
\cite{Morris:2016spn,Percacci:2016arh,Ohta:2017dsq,Nieto:2017ddk}.
The physical meaning is very different, though:
the split transformation is the freedom of shifting
the background and the quantum field by equal and opposite amounts
and is always an invariance of the classical action.
The cutoff, however, introduces separate dependences on these two variables and breaks the split transformations.
For transformations of the background metric
of the form (\ref{scaletr}), with constant $\epsilon$,
the anomalous WI contains the term 
$\epsilon\partial_t\Gamma_k$ in the r.h.s.
The difference with the physical scale transformations
considered in this paper is highlighted by the 
invariance of the measure under split scale transformations,
which results in the absence of the term $-\cA$.

\medskip

\noindent
{\bf Acknowledgements}
R.P. would like to thank G.P. Vacca and J. Pawlowski for
discussions in the early stages of this work. T.R.M. acknowledges support from both the Leverhulme Trust and the Royal Society as a Royal Society Leverhulme Trust Senior Research Fellow, from STFC through Consolidated Grant ST/P000711/1, and thanks SISSA for support under the Collaborations of Excellence program.

\appendix{}

\section{Rescaling the coordinates vs. rescaling the metric}

A scale transformation is a change of all lengths by a common factor.
Since physical lengths are defined by integrating the line element
$ds^2=g_{\mu\nu}dx^\mu dx^\nu$,
a scale transformation can be interpreted either as a scaling of the
metric or as a scaling of the coordinates.
In the main text we have followed the former convention,
which is more natural from the point of view of General Relativity.
In flat space QFT it is customary to define the scale transformations
as rescalings of the coordinates: 
$\delta_\epsilon x^\mu=\epsilon x^\mu$.
Then the infinitesimal transformation of the fields is 
$\delta_\epsilon\phi=\epsilon(-x^\mu\partial_\mu+d_\phi)\phi$.
The canonical dimension of a field, $d_\phi$,
which is determined by requiring scale invariance
of the kinetic term, is the same in both cases.
The canonical energy-momentum tensor comes from the Noether
current associated to translation invariance.
In general it does not coincide with the one defined in (\ref{emt}),
but there are well-known ``improvement'' procedures that make
them equal.

Equations (\ref{q2},\ref{intq},\ref{laptrans}) hold also 
when one rescales the coordinates, and so the derivation
of the scale WI in section 3.1 goes through in the same way.
The mass dimension of an operator $\cO$, 
containing $n_\phi$ fields and $n_\partial$ derivatives is
$\Delta=-d+n_\partial+\frac{d-2}{2}n_\phi$,
where $-d$ comes from $d^dx$.
This is equal to the expression given in (\ref{delta}),
because $n_g$ is $d/2$, coming from $\sqrt{g}$,
minus the number of inverse metrics, which must be equal
to half the number of derivatives.


\end{document}